\documentclass[prb,aps,showpacs,onecolumn,superscriptaddress,floatfix]{revtex4-1}

\usepackage{epsfig}
\usepackage{amsmath}
\usepackage{amssymb}
\usepackage{amsfonts}
\usepackage{mathptmx}
\usepackage{eucal}
\usepackage{bm,bbm}
\usepackage{color}
\usepackage{bbold}

\usepackage{epstopdf}

\usepackage{graphicx}

\usepackage{natbib}

\def\rv{{\bf r}}

\def\nv{{\bf n}}
\def\pv{{\bf p}}

\def\xiv{\bm{\xi}}

\def\Fcal{{\mathcal F}}
\def\Acal{{\mathcal A}}
\def\tr{{\rm tr}}
\def\fpar{f_t^{\parallel}}
\def\fperp{{\bf f}_t^{\perp}}

\def\Gbb{\mathbbm{G}}
\def\gbb{\mathbbm{g}}
\def\Sigmabb{\mathbb{\Sigma}}

\newcommand{\sgn}{{\mathrm{sgn}}}

\begin{document}
\title{Spin-orbit coupling as a source of long-range triplet proximity effect in superconductor-ferromagnet hybrid structures}
\author{F. S. Bergeret}
\affiliation{Centro de F\'{i}sica de Materiales (CFM-MPC), Centro
Mixto CSIC-UPV/EHU, Manuel de Lardizabal 4, E-20018 San
Sebasti\'{a}n, Spain}
\affiliation{Donostia International Physics Center (DIPC), Manuel
de Lardizabal 5, E-20018 San Sebasti\'{a}n, Spain}
\author{I. V. Tokatly}
\affiliation{Nano-Bio Spectroscopy group, Dpto. F\'isica de Materiales, Universidad del Pa\'is Vasco, Av. Tolosa 72, E-20018 San Sebasti\'an, Spain}
\affiliation{IKERBASQUE, Basque Foundation for Science, E-48011 Bilbao, Spain}

\date{\today }

\begin{abstract}
We investigate the proximity effect in diffusive superconducting hybrid structures with a spin-orbit (SO) coupling. Our study is focused on the singlet-triplet conversion and the generation of long-range superconducting correlations in ferromagnetic elements. We derive the quasiclassical equations for the Green's functions including  the SO coupling terms in form of a background SU(2) field. With the help of these equations, we first present an interesting complete analogy between the spin diffusion process in normal metals and the generation of the triplet components of the condensate in a diffusive superconducting structure in the presence of  SO coupling. From this analogy it turns out naturally that the SO coupling is an additional source of the long-range triplet component (LRTC)  besides the magnetic inhomogeneities studied in the past.  This  analogy opens a range of possibilities for the generation and manipulation of the triplet condensate in hybrid structures. In particular we demonstrate that a normal 
metal with a SO coupling can be used as source of LRTC if attached to  a S/F bilayer. We also  demonstrate an explicit connection between an inhomogeneous exchange field and SO coupling mechanisms for the  generation of the  LRTC and  establish the conditions for the appearance of the LRTC in different geometries.  Our work gives a global description of the singlet-triplet conversion in hybrids structures in terms of generic spin-fields and our results are particularly important for the understanding of the physics underlying spintronic devices with superconductors.
\end{abstract}
 
\maketitle

\section{Introduction}

It is by now common knowledge that the interaction between conventional superconductivity and ferromagnetism in superconductor-ferromagnet (S/F) hybrids leads to a new type of superconducting correlations in a triplet state\cite{Bergeret2001,BergeretRMP}.  Since the prediction of this intriguing phenomenon in 2001, there has been an increasing experimental activity in the field.\cite{Keizer2006,Birge2010,Robinson2010, Robinson2010b,Leksin2012,Witt2012,Wang2012,Robinson2012,Chiodi,Pal2014,Robinson2014,Kalchemin,Banerjee,Aarts2010, Wang2010, Sprungmann2010, Usman2011,Birge2012, Villegas2012, Aarts2012,Yates2013} That research  focuses mainly on the creation and control of superconducting triplet correlations in hybrid structures with the ultimate goal of using polarized spin supercurrents in spintronic devices.\cite{EschrigPT} To achieve this,  it is essential to  identify the optimal material combination and hence it is of fundamental interest  to understand the  physics that underpin triplet generation.

In  S/F structures, superconducting correlations can penetrate into the ferromagnetic metal due to the  proximity effect. If the ferromagnet is a monodomain magnet, the superconducting condensate consists of two components: the usual singlet one and the triplet component with total zero spin-projection with respect to the magnetization axis of the F layer.  In a diffusive system both components decay over a short distance given by $\sqrt{D/h}$, where $D$ is the diffusion coefficient of the F layer and $h$ the exchange field.  If, however, the triplet components with finite total spin are generated, these can penetrate the F region over a much longer distance of the order of $\sqrt{D/2\pi T}$.
 
It is commonly believed that singlet-long-range triplet (in short singlet-triplet) conversion happens only in the presence of magnetic inhomogeneities, as for example magnetic  domain walls,\cite{Bergeret2001,Fominov2007,Volkov2008,Kupferschmidt2009, Buzdin2011} ferromagnetic multilayers with different magnetization orientations\cite{Bergeret2003,Houzet2007,Halterman2008}, or  interfaces with magnetic disorder.\cite{Eschrig2008,Linder2009}
Such inhomogeneities presumably explain the observation of long-ranged Josephson currents through Ho-Co-Ho bridges, due to the spiral-like magnetization of the Ho layers\cite{Robinson2010}, or though ferromagnetic X-Co-X multilayers, where  the inhomogeneous magnetization of X=PdNi,CuNi, Ni might act as spin-mixers\cite{Birge2010,Birge2012}
More surprising  is the observation of a long-range Josephson effect  in  lateral   structures based on  the half-metallic  CrO$_2$.\cite{Keizer2006,Aarts2010,Aarts2012} A first explanation for such observations assumes a spin-active interface between the CrO$_2$ layer and the superconductor, consequence of a magnetic inhomogeneity at the atomic level.\cite{Eschrig2008}

In ballistic heterostructures it has been shown  that spin-orbit (SO) coupling can also be a source for a triplet superconducting condensate.\cite{Edelstein2003,Duckheim2011,Takei2012,Takei2013,jain0} Being  anisotropic in momentum  this condensate component is very sensitive to disorder and vanishes in diffusive systems.    However, in a recent work we have demonstrate that  in  S/F diffusive systems a finite spin-orbit (SO) coupling can be also a source for the s-wave long-range triplet correlations (LRTC).\cite{BT2013}  A finite SO coupling can  result from either  an intrinsic property of materials without inversion symmetry\cite{Samokhin2009} or from  geometrical constraints such as low dimensional structures or  interfaces between different materials.\cite{Rashba,Edelstein2003,Koroteev2004,Ast2007,Miron2010,Linder2011,Duckheim2011,Takei2012,Takei2013}  Specifically, Ref.~\cite{BT2013} presented a unified view of the singlet-triplet conversion which connects the magnetic inhomogeneous mechanism with the 
one based on SO coupling.

In the  present work we readdress the problem of  singlet-triplet conversion in diffusive S/F structures in the presence of arbitrary (linear in momentum) spin-orbit coupling and go a step further. The main goal of the present paper is twofold: First,  we present a complete analogy between the diffusion of a spin density  in a normal metal and the singlet-triplet conversion in superconducting hybrids. This analogy opens a new view of the singlet-triplet conversion that helps in the understanding of the proximity effect in more complex hybrid structures.  Second,  we present the derivation of quasiclassical equations in the presence of a SO-coupling and superconducting correlations. These equations can be very useful not only to describe the singlet-triplet conversion but also for the study of the dynamics of S/F hybrids.  With the help of these equations we analyze different hybrid structures and discuss the condition for the singlet-triplet conversion. In particular we show that all triplet components can 
be generated in a S/F/N structure, provided the conductor N exhibits a SO coupling. We also show that while for a transverse multilayer structure of S/F/S type, the "old" picture of magnetic inhomogeneities can explain  the long-range Josephson coupling\cite{Birge2010,Robinson2010}, in lateral S/F structures the SO mechanism may be consider as the main mechanism for singlet-triplet conversion.\cite{Keizer2006,Aarts2010}

The structure of the paper is the following: In the next section we review the spin diffusion in the normal case. We discuss the  spin diffusion and  relaxation  in a normal metal in the presence of a generic SO coupling, placing emphasis on the  main mechanism that can change the direction of the spin. In section \ref{sec-diffS-SO} we discuss the singlet-triplet conversion in a proximity metal with SO coupling and draw  an analogy between the singlet-triplet conversion and the "precession" of the spin density in the normal state. In section \ref{sec-diffS-Bloch} we readdress the original  problem of singlet-triplet conversion in a Bloch domain wall\cite{Bergeret2001} and show that it is gauge-equivalent to the one of a ferromagnet with a homogeneous exchange field and SO coupling. In the previously mentioned sections  we base our  analysis on a heuristic SU(2) covariant diffusion equation. A  rigorous derivation of the quasiclassical kinetic equation for the Green function is  presented in section \ref{sec-equations}. We present both non-equilibrium (Keldysh), and equilibrium (Matsubara) formalisms. In section \ref{sec-examples} we discuss hybrid structures of different geometries. We show that the triplet component with a finite total spin can be generated in a S/F/N structure with a homogeneous magnetized F, provided SO coupling in the  N metal.  We also analyze a transversal and longitudinal S/F structure and demonstrate  that even in the case of a homogeneous magnetization, an interfacial SO coupling can generate long-rang correlations.  Finally we  present some discussions and a summary of results  in our concluding section.

\section{Spin diffusion and relaxation in normal systems with spin-orbit coupling}
\label{sec-diffN}

To understand how SO coupling can affect the proximity effect in S/F systems, 
it is  instructive to recall the physics of  spin diffusion in a normal system.
For this sake we consider a normal conductor  described by the Hamiltonian
 \begin{equation}
   \label{H-generalSO}
 H_0 = \frac{{\bf p}^2 }{2m} - \frac{1}{2}\Omega^{a}({\bf p})\sigma^{a} + V_{imp}  
 \end{equation}
where $V_{imp}$ is the spin-independent potential of randomly distributed impurities, and the second term, $H_{SO}=\frac{1}{2}\Omega^{a}({\bf p})\sigma^{a}$ with $\Omega^{a}({-\bf p})=-\Omega^{a}({\bf p})$, describes a generic SO coupling allowed in any system with lack of inversion symmetry. The matrices $\sigma^a$, with $a=x,y,z$, are the Pauli matrices. Physically, the above SO coupling corresponds to an effective momentum-dependent Zeeman field which induces precession of the electron spin about the direction of the vector ${\bm \Omega}({\bf p}) = \{\Omega^{x}({\bf p}),\Omega^{y}({\bf p}),\Omega^{z}({\bf p})\}$.

In this work we consider spin dynamics in the diffusive limit, {\it i.~e.} when the elastic mean free path $l=\tau v_{F}$ (here $\tau$ is the momentum relaxation time and $v_{F}$ is the Fermi velocity) is much shorter then the  other length scales. In this limit the spin density vector ${\bf S}=(S^x,S^y,S^z)$ obeys the spin diffusion equation presented in Eq.~(\ref{diffusion1}) below. To make our argumentation self-contained we give a general and compact symmetry-based derivation of this equation. 

To reveal the structure of the spin diffusion equation in such systems it is instructive to consider a special, but still rather general type of linear in momentum SO coupling with 
\begin{equation}
\Omega^{a}({\bf p})={\Acal}_{k}^{a}\frac{p_k}{m}. 
\label{linearSO}
\end{equation}
The mathematical beauty of the linear coupling is related to a local SU(2) gauge invariance of the corresponding Hamiltonian\cite{Mineev92,Frolich93,Jin2006,Tokatly2008} that can be written (up to an irrelevant constant) as follows
 \begin{equation}
\label{H-linearSO}
H_0 = \frac{1}{2m}(p_j - \hat{\Acal}_j)^2 +V_{imp}, 
\end{equation}
where  $\hat{\Acal}_j=\frac{1}{2}\Acal_j^a\sigma^a$. The first term in  Eq.~(\ref{H-linearSO}) formally describes nonrelativistic particles minimally coupled to a $2\times 2$ matrix-valued SU(2) vector potential $\hat{\Acal}_j$. Hence the SO coupling enters the problem as an effective background SU(2) field.  This implies the form-invariance of  the Hamiltonian (\ref{H-linearSO}) under any local SU(2) rotation with a matrix $\hat{U}=e^{\frac{i}{2}\chi^a(\rv)\sigma^a}$ supplemented with the gauge transformation of the potential $\hat{\Acal}_j\mapsto\hat{U}\hat{\Acal}_j\hat{U}^{-1} -i (\partial_j\hat{U})\hat{U}^{-1}$. Many general aspects of spin physics in SO coupled systems acquire a simple interpretation in terms of this gauge invariance\cite{Tokatly2008,Yang06,Liu07,Natano07,Yang08,Tokatly_Ann,Tokatly10,Gorini2010}.

In the diffusive limit, for systems without SO coupling the spin-density matrix $\hat \rho$ obeys the standard diffusion equation: $\partial_t\hat \rho=D\nabla^2\hat \rho$, where $D$ is the diffusion constant. If the SO coupling  is present, the gauge invariance arguments tell us that all we need is to replace the derivatives by their covariant counterparts, i.~e. $\partial_{k}\cdot\mapsto \tilde{\nabla}_k\cdot=\partial_k\cdot-i[\hat{\Acal}_k,\cdot]$. This replacement ensures that the spin-density matrix transforms covariantly, $\hat \rho\mapsto\hat{U}\hat \rho\hat{U}^{-1}$, under a local SU(2) rotation. Therefore the spin diffusion equation takes the form
\begin{equation}
\label{diffusion0}
\partial_t\hat \rho=D\tilde{\nabla}^2\hat\rho\; ,
\end{equation}
where the right hand side of this equation encodes the effects of the SO coupling in the diffusive regime.
For a spatially uniform SO field the covariant Laplacian $\tilde{\nabla}^2$ in Eq.~(\ref{diffusion0}) reads 
\begin{equation}
\label{Laplace}
\tilde{\nabla}^2\hat\rho=\nabla^2\hat\rho - 2i[\hat{\Acal}_k,\partial_k\hat\rho] - [\hat{\Acal}_j,[\hat{\Acal}_j,\hat\rho]]
\end{equation}
The physical significance of the last two, SO induced,  terms in Eq.~(\ref{Laplace}) becomes more clear if we rewrite the spin diffusion equation (\ref{diffusion0}) in terms of the spin density vector with components $S^a =\frac{1}{2}\tr\{\hat\rho\sigma^a\}$
\begin{equation}
 \label{diffusion1}
 \partial_t S^a = D\nabla^2 S^a + 2 C_k^{ab}\partial_k S^b - \Gamma^{ab}S^b,
\end{equation}
where the tensors $ C_k^{ab}$ and $\Gamma^{ab}$ are defined as follows
\begin{eqnarray}
 \label{C-linear}
 C_k^{ab} &=& D\varepsilon^{acb}\Acal_k^c, \\
 \label{Gamma-linear}
 \Gamma^{ab} &=& D\left(\Acal_k^c\Acal_k^c\delta^{ab} - \Acal_k^a\Acal_k^b\right),
\end{eqnarray}
and $\varepsilon^{acb}$ is the Levi-Civita tensor.
The symmetric, positive semidefinite tensor, $\Gamma^{ab}\equiv (1/\tau_s)^{ab}$  in Eq.~(\ref{diffusion1}) originates from the double commutator in Eq.~(\ref{Laplace}) and  is responsible for the (anisotropic) Dyakonov-Perel  spin relaxation.\cite{DP_1971a,DP_1971b} 
The second term in the right hand side of Eq.~(\ref{diffusion1}) describes the  precession of  the spin of diffusively moving  particles in the presence of a spatially  inhomogeneous spin distribution.  

It is worth outlining that by considering a seemingly special, linear in momentum SO coupling and using only the gauge invariance requirements we actually recovered the most general form Eq.~(\ref{diffusion1}) of the spin diffusion equation (see e.~g. Ref.~\cite{Stanescu2007,Yang2010}). In fact, the formal quantum kinetic derivation of the spin diffusion equation for the most general SO coupling with arbitrary ${\bm\Omega}({\bf p})$ yields Eq.~(\ref{diffusion1}). The only difference is that now the tensors $ C_k^{ab}$ and $\Gamma^{ab}$ are defined by more general, but structurally similar to Eqs.~(\ref{C-linear})-(\ref{Gamma-linear}), expressions:
\begin{eqnarray}
 \label{C-general}
 C_k^{ab} &=& \tau_{p}\varepsilon^{acb}\langle v_k(\pv)\Omega^c(\pv)\rangle_{F}, \\
 \label{Gamma-general}
 \Gamma^{ab} &=& \tau_{p}\langle\Omega^c(\pv)\Omega^c(\pv)\delta^{ab} - \Omega^a(\pv)\Omega^b(\pv)\rangle_{F}\; ,
\end{eqnarray}
where $v_k(\pv)=\frac{\partial \varepsilon(\pv)}{\partial p_k}$ is the $k$-component of the particle velocity, and $\langle\dots\rangle_{F}$ stands for the Fermi surface averaging. The important conclusion is that most qualitative physical results (at least in the diffusive regime) obtained for the linear SO coupling should be valid generically for any noncentrosymmetric system.

We now  discuss the main features of the spin diffusion, which will be relevant for the problem of singlet-triplet conversion in  superconducting  hybrid structures. We consider for simplicity a system with one-dimensional inhomogeneity along the x-axis and  assume that by injecting a spin current at $x=0$  one  creates a finite $z$-component $S_0^z$ of the spin density at the origin.  The injected spin diffuses into the system according to Eq. (\ref{diffusion1}). We now analyze the resulting stationary spatial distribution of the spin density ({\it i.e} $\partial_t S^a =0$)  by solving the stationary 1D version of Eq.~(\ref{diffusion1}),
\begin{equation}
 \label{diffusion3}
D\partial_x^2 S^a + 2 C_x^{ab}\partial_x S^b - \Gamma^{ab}S^b = 0,
\end{equation}
with the boundary condition ${\bf S}(x=0)=\hat{\bf z} S_0^z$. Beside the  decay  away from $x=0$ due to the Dyakonov-Perel spin relaxation, the two last terms in Eq.~(\ref{diffusion3}) encode two possible mechanisms of the spin rotation in the presence of SO coupling.

\begin{figure}[t]
\begin{center}
\includegraphics[width=\columnwidth]{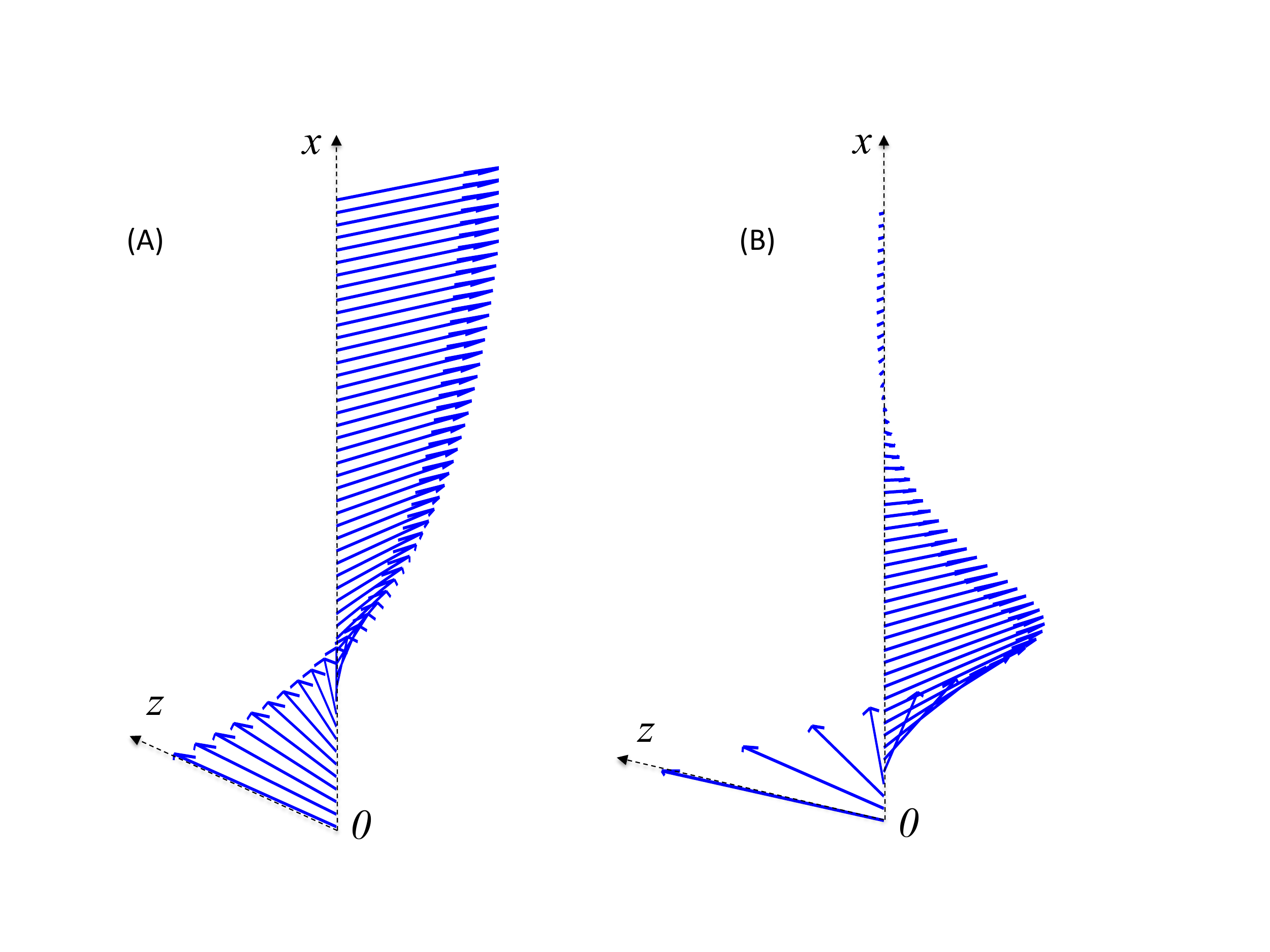}
\caption{\label{fig-spiral}Schematic view of the spin rotation in a layered system with 1D inhomogeneity. At $x=0$ a spin parallel to the $z$-axis is injected. Due to the SO coupling also the $S^y$ component becomes finite upon diffusion. Panel (A)  illustrates the spin rotation due to anisotropy of the Dyakonov-Perel tensor $\Gamma$.  The vectors are given by Eqs.(\ref{rot1}-\ref{rot2}) and  we have chosen $\beta/\alpha=1/3$.  Panel (B) shows the spin rotation due to the second term in Eq. (\ref{diffusion3})  for an isotropic SO coupling.  The vectors are given in Eqs. (\ref{helix-Sz}-\ref{helix-Sy}) .}
\end{center}
\end{figure}

The first mechanism is related to the fact that  the spin relaxation tensor $\Gamma^{ab}$ in general can be anisotropic. This means that different components of the spin may have different relaxation rate. If it happens that the injected spin is not parallel to one of the principal axes of $\Gamma^{ab}$, the spin will rotate in the course of diffusion by turning towards the direction with the slowest relaxation rate.  In order to illustrate the evolution of the spin  due to this mechanism we assume that the SO coupling  is described by $\Acal^x_z=\beta$, $\Acal^y_z=\alpha$ and the rest of the components of the $\Acal^a_k$ tensor equals to zero. In such a case the second term of Eq. (\ref{diffusion3}) vanishes and the solution with ${\bf S}(x=0)=\hat{\bf z} S_0^z$ is given by:
\begin{eqnarray}
\frac{S^z(x)}{S^z_0}&=&\frac{\beta^2}{\alpha^2+\beta^2}+\frac{\alpha^2}{\alpha^2+\beta^2}e^{-\kappa x}\label{rot1}\\
\frac{S^y(x)}{S^z_0}&=&\frac{\alpha\beta}{\alpha^2+\beta^2}-\frac{\alpha\beta}{\alpha^2+\beta^2}e^{-\kappa x}\label{rot2}\; ,
\end{eqnarray}
where $\kappa=\sqrt{\alpha^2+\beta^2}$. In Fig.\ref{fig-spiral}A we sketched  the spatial evolution of the spin. One clearly sees that the injected spin, originally parallel to the z axis, rotates and acquires a finite $y$ component due to the SO coupling.

%%%%%%%%%%%%%%%%%%%%%%%%%%%%%%%%%%%%%%%%%%%%%%%%%%%%%%%%%%%%%%%%%%%
%%%%%%%%%%%%%%%%%%%%%%%%%
%%%%%%%%%%%%%%%%%%%%%%%%%%%%%%%%%%%%%%%%%%
%%%%%%%%%%%%%%%%%%%%%%%%%%%%%%%%%%%%%%%%%%%%%%%%%%%%%%%%%%%%%%%%%%%

The second mechanism for spin rotation  is the  ``precession'' generated by the second term in the right hand side of  Eq.~(\ref{diffusion3}). This mechanism is operative even for systems with equal relaxation rates for all spin directions.
To illustrate the effect of this term we consider the simplest fully isotropic SO coupling described by  the diagonal SO field $\Acal^a_j=\alpha \delta_{j}^{a}$. In this case Eq.~(\ref{diffusion3}) reduces to the following system of coupled diffusion equations for the spin components $S^z(x)$ and and $S^y(x)$
\begin{eqnarray}
\label{diffusion-Sz}
D\partial_{x}^2 S^z + 2D\alpha\partial_x S^y - \frac{S^z}{\tau_{s}} &=&0,\\
\label{diffusion-Sy}
D\partial_{x}^2 S^y - 2D\alpha\partial_x S^z - \frac{S^y}{\tau_{s}}&=&0,
\end{eqnarray}
where $\tau_{s}=1/(2D\alpha^2)$ is (now isotropic) spin relaxation time [in deriving this equations we made use of Eqs.~(\ref{C-linear}) and (\ref{Gamma-linear})]. The coupling of different components of the spin in Eqs.~(\ref{diffusion-Sz})-(\ref{diffusion-Sy}) has a typical precession structure -- it induces precession of the spin direction around the direction of inhomogeneity. Straightforward solution of these equations with the boundary condition  ${\bf S}(x=0)=\hat{\bf z} S_0$ yields the helicoidal spin distribution,
\begin{eqnarray}
 \label{helix-Sz}
 S^z(x) = S^z_0 e^{-\alpha x}\cos{\alpha x}, \\
 S^y(x) = S^z_0 e^{-\alpha x}\sin{\alpha x},
 \label{helix-Sy}
\end{eqnarray}
which clearly demonstrates the effect of the precession term in the spin diffusion equation. The injected spin relaxes and rotates provided there is a spatial component of the SO field $\Acal_k^a$, or, more generally, the tensor $C_k^{ab}$,  along the direction of inhomogeneity. In Fig.~\ref{fig-spiral}B we show schematically the spin rotation described by Eqs.~(\ref{helix-Sz}-\ref{helix-Sy}). 

In short, there are two mechanisms that can change the direction of the injected spin density. One originates from a possible anisotropy of the spin relaxation rate tensor $\Gamma^{ab}$ in Eq. (\ref{diffusion1}), while the other mechanism is due to precession of the spin when $\hat\rho$ is spatially inhomogeneous according to the second term in the left hand side of Eq.~(\ref{diffusion1}). In the next section we show  that these well established mechanisms for rotation of the spin also explain the rotation of the triplet component of the superconducting condensate and the appearance of a long-range proximity effect in SF hybrid structures with SO coupling.

\section{The singlet-triplet conversion in diffusive S/F structures: a physical picture}
\label{sec-diffS}

\subsection{LRTC in S/F structures with SO coupling}
\label{sec-diffS-SO}
We now discuss the singlet-triplet conversion in S/F structures in the presence of SO coupling. To pursue our line of reasoning we present  in this section the linearized equation that describes the proximity effect in S/F structures and postpone its derivation to the next section.

\begin{figure}[h]
\begin{center}
\includegraphics[width=\columnwidth]{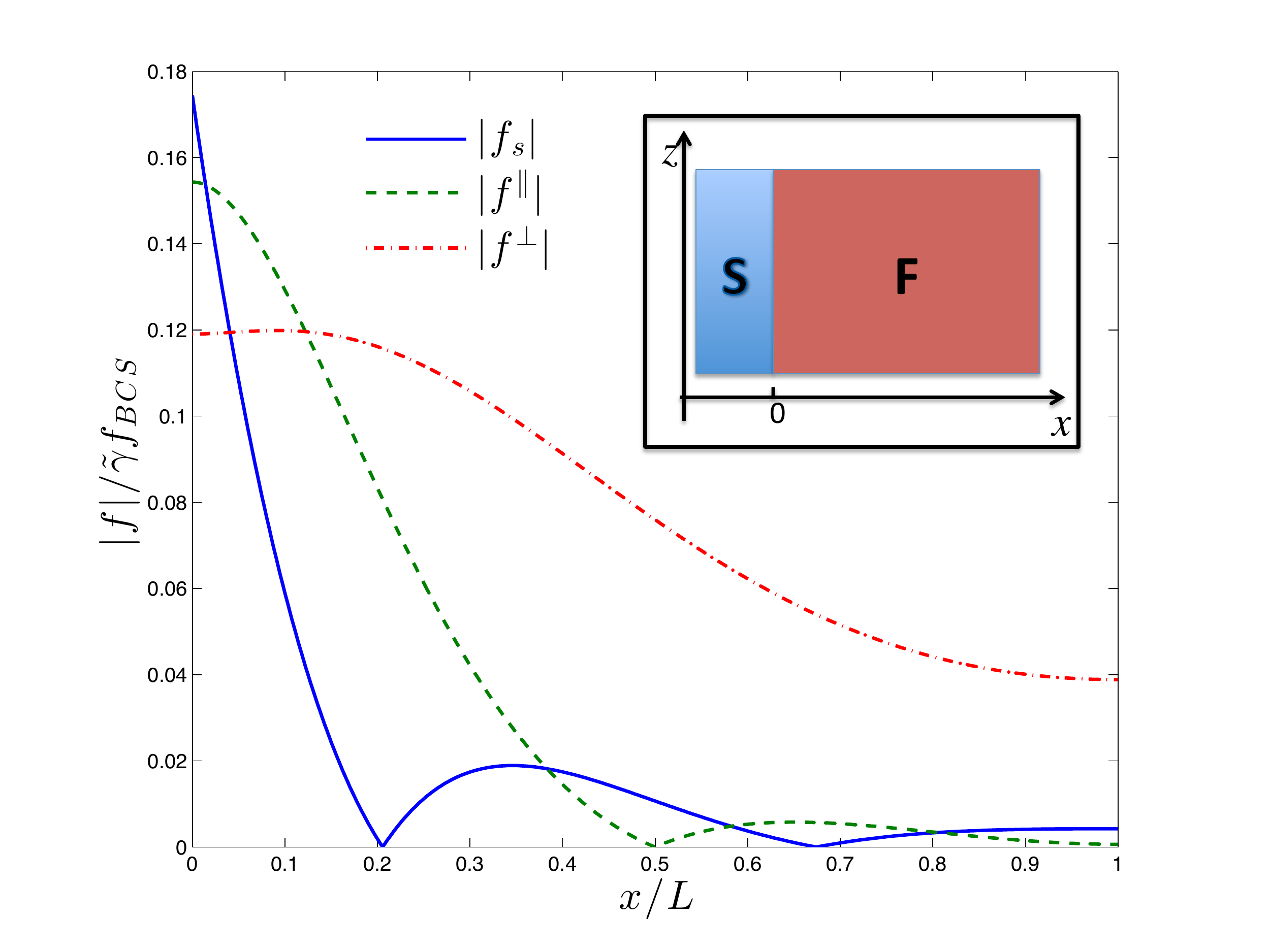}
\caption{\label{fig_sol} The spatial dependence of the amplitude of all components of the condensate function for the geometry shown in the inset and obtained from Eqs. (\ref{Usadel-s-example}-\ref{Usadel-perp-example}) . The exchange field in F is homogeneous and points in $z$ direction. A fully isotropic SO coupling is assumed in F.    We have chosen $h=10\Delta$, $\omega=\pi T$, $T=0.1\Delta$,  $\alpha\xi_0=1$ and $L=1.5\xi_0$. Here $\Delta$ is the superconducting gap in S and the length $\xi_0$ is defined as $\xi_0=\sqrt{D/\Delta}$. }
\end{center}
\end{figure}

We consider first the proximity effect in S/F structures without SO coupling. For simplicity we assume that the proximity effect is weak and therefore our starting point is the  linearized Usadel equation\cite{Usadel} which describes  the superconducting condensate $\hat f$ induced  in the  diffusive ferromagnet F (see  inset in Fig.~\ref{fig_sol}) 
 \begin{equation}
 \label{Usadel-gen}
D\nabla^2\hat{f} - 2|\omega|\hat{f} -i\sgn(\omega) \{\hat{h},\hat{f}\}=0\; .
\end{equation}
Here  $\omega$ is the Matsubara frequency and  $\hat{h}=h^a\sigma^a$ is the exchange field whose vector components $h^a$ may depend on space coordinates. This well-known equation, which has been used in most previous works on S/F structures (see for example Ref.~\cite{BergeretRMP} and references therein), describes diffusion of the condensate in the ferromagnet. The generation (injection) of the $s$-wave condensate at the S/F interface is commonly described by the Kupriyanov-Lukichev  boundary condition \cite{KL} which in its linearized version has a simple form
\begin{equation}
 \label{BC-gen}
N_{k}{\partial}_{k}\hat{f}\big\vert_{S/F} = -\gamma f_{BCS},
\end{equation}
where $f_{BCS}=\Delta/\sqrt{\omega^2+\Delta^2}$ is the anomalous Green's function in the S-region, $N_k$ the $k$-component of the  vector normal to the S/F interface, and $\gamma$ is a parameter that describes the quality of the S/F barrier. The boundary condition (\ref{BC-gen}) works for interfaces with low transmission while  for a perfect transparent barrier one should impose the continuity of $\hat f$ at the S/F interface.

Let us briefly recall the widely studied proximity effect in S/F structures without SO coupling. 
The most general form of the condensate function $\hat f$  satisfying Eqs.~(\ref{Usadel-gen})-(\ref{BC-gen})  is:\cite{Champel2005,Fominov2006}
\begin{equation}
\label{condensate}
\hat f=f_s\hat 1+f_t^a \sigma^a\; .
\end{equation}
Here $f_s$ is the singlet component which is scalar in the spin space, while ${\bf f}_t$ is a vector in spin space (with components $f_t^a$) describing the triplet component. In the case of a spatially homogeneous exchange field ${\bf h}$ the condensate induced in F-region acquires both the singlet component $f_s$ and the triplet component ${\bf f}_t = \fpar {\bf h}/{h}$ with the spin along ${\bf h}$. Because of these two components the anticommutator in the right hand side of  Eq.~(\ref{Usadel-gen}) is nonzero thus providing a coupling between the singlet $f_s$ and the parallel to ${\bf h}$ triplet $f^\parallel$ condensates. The magnitude of the coupling is given by the amplitude $h$ of the exchange field that is typically much larger than the characteristic energy ($\sim T$) of  the second term in Eq.~(\ref{Usadel-gen}).  Thus, the  decaying length for both components away from the S/F interface is controlled by the  singlet-triplet coupling, being of the order of $\xi_h=\sqrt{D/h}$. In other words, in the 
presence 
of a large  exchange field the proximity effect becomes short ranged. 

The structure of Eq.~(\ref{Usadel-gen}) suggests a way to circumvent the fast decay of superconducting correlations in ferromagnets. If by some means we generate components of the triplet condensate in any direction perpendicular to ${\bf h}$ the anticommutator in Eq.~(\ref{Usadel-gen}) vanishes and therefore those perpendicular components $\fperp$ will decay over the scale of the order of $\sqrt{D/T}$ which is much larger than $\xi_h$. It is very well established  that such a long-range component, $\fperp$, can be induced in the presence of a spatially inhomogeneous vector ${\bf h}$\cite{Bergeret2001,BergeretRMP}. But only  recently it has been shown that SO coupling provides   an alternative mechanism for
 generating the long-range triplet condensate.\cite{BT2013}

Physically generation of the perpendicular component $\fperp$ can be viewed as a rotation of the triplet pair spin away from the direction of the exchange field.\cite{EschrigPT} In the previous section we have seen that such a rotation is a generic feature of the spin diffusion in the presence of SO coupling and, as we now show,  this feature should not depend on the nature of spin carriers, whether they are single electrons or triplet Cooper pairs.

In the presence of SO coupling the Usadel equation  should be properly modified. As  previously done, we consider only  linear in momentum SO coupling describe by the  Hamiltonian (\ref{H-linearSO}). In complete analogy with the spin diffusion in a normal system (see Section~\ref{sec-diffN}) the SO-coupling-modified Usadel equation is obtained from Eq.~(\ref{Usadel-gen}) by replacing all derivatives with their covariant counterparts, 
$\partial_{k}\cdot\mapsto \tilde{\nabla}_k\cdot=\partial_k\cdot-i[\hat{\Acal}_k,\cdot]$,
\begin{equation}
 \label{cov-Usadel-gen}
D\tilde{\nabla}^2\hat{f} - 2|\omega|\hat{f} -i\sgn(\omega) \{\hat{h},\hat{f}\}=0\; .
\end{equation}
To ensure that the condensate function $\hat{f}$ is transformed covariantly under a local SU(2) rotation the Kupriyanov-Lukichev boundary condition (\ref{BC-gen}) should be also modified accordingly,
\begin{equation}
 \label{cov-BC-gen}
N_{k}\tilde{\nabla}_{k}\hat{f}\big\vert_{S/F} = -\gamma f_{BCS}.
\end{equation}
The system of Eqs.~(\ref{cov-Usadel-gen}), (\ref{cov-BC-gen}) describes the spatial distribution of the superconducting condensate induced from a s-wave superconductor in a ferromagnet with SO coupling.  The covariant derivatives in these equations encode again all effects of SO coupling. If we now substitute  the representation of Eq.~(\ref{condensate}) for the condensate function we obtain
\begin{eqnarray}
 \label{cov-Usadel-gen-s}
&&D{\nabla}^2f_s - 2|\omega|f_s - 2i\sgn(\omega) h^a f_t^a=0\; , \\
\label{cov-Usadel-gen-t}
&&D\nabla^2 f_t^a + 2 C_k^{ab}\partial_k f_t^b - \Gamma^{ab}f_t^b- 2|\omega|f_t^a - 2i\sgn(\omega) h^a f_s = 0\; ,
\end{eqnarray}
 from   Eq.~(\ref{cov-Usadel-gen}), and 
\begin{eqnarray}
\label{cov-BC-gen-s}
&&N_{k}\partial_{k}f_s\big\vert_{S/F} = -\gamma f_{BCS}\; ,\\
 \label{cov-BC-gen-t}
&&N_{k}(\partial_{k}f_t^a + C_k^{ab} f_t^b)\big\vert_{S/F} = 0\; ,
\end{eqnarray}
from the  boundary condition of Eq.~(\ref{cov-BC-gen}).
We have used the definitions of  the Dyakonov-Perel spin relaxation tensor $\Gamma^{ab}$ and the diffusive spin precession tensor $C_k^{ab}$ presented in  Eqs.~(\ref{C-linear}) and (\ref{Gamma-linear}). In the most general SO coupling, one can show, that  the structure of Eqs.~(\ref{cov-Usadel-gen-s}-\ref{cov-BC-gen-t}) remains the same with the tensors  $\Gamma^{ab}$ and $C_k^{ab}$ redefined according to Eqs.~(\ref{C-general}) and (\ref{Gamma-general}).

The comparison of Eq.  (\ref{cov-Usadel-gen-t}) with  the spin diffusion equation (\ref{diffusion1}) shows  the  complete analogy  between  spin diffusion in normal  and superconducting systems.  In particular, the physical effect of SO coupling is practically identical to that discussed in Sec.~\ref{sec-diffN}. 
 
By inspection of Eq.  (\ref{cov-Usadel-gen-t}), it becomes clear that  the direction of the condensate spin is not preserved in the F-region, due to the  SO coupling.  Similarly to the normal case the second and the third terms in Eq.~(\ref{cov-Usadel-gen-t}) describe two mechanisms of the spin rotation -- (i) a possible anisotropy of the relaxation rate, and (ii) the spin precession in the presence of a spatially inhomogeneous spin density. Therefore in the course of diffusion the spin of the condensate turns away from the direction of the exchange field. In other words a  component  perpendicular to ${\bf h}$  appears and  decays over a length scale much larger than $\xi_h$. This slowly decaying part of ${\bf f}_t$  is responsible for the long-range proximity effect in S/F structures.

It is worth noting that the anisotropy of the spin relaxation rate generates the LRTC only if the direction of the exchange field does not coincide with one of the principal axes of the relaxation rate tensor $\hat\Gamma$. However, it is natural to expect that in realistic ferromagnets both ${\bf h}$ and the principal axes of $\Gamma^{ab}$ are linked to some crystallographic directions. Therefore it is quite probable that they do coincide and the mechanism (i) along is not sufficient to induce the LRTC in most of realistic situations. The second mechanism (ii), {\it i.e.}  the spin precession mechanism is more likely to occur and more universal. 
% The only restriction is that the effective SO field should have a spatial component along the direction of the spin inhomogeneity.  

Because of its practical importance it is useful to have a simple illustrative example for the  generation of LRTC via the spin precession mechanism. Let us consider the structure sketched in the inset of Fig.\ref{fig_sol}. It is a S/F structure with the interface perpendicular to the $x$-axis (${\bf N}=\hat{\bf x}$)  and  the exchange field ${\bf h}=\hat{\bf z}h$ along $z$-axis. We assume a fully isotropic SO coupling with $\Acal_k^a=\alpha\delta_k^a$.  By assuming that the structure has infinite dimensions in the z-y plane, the condensate function is invariant in this directions and only depends on x. Moreover, by   symmetry, the triplet condensate function ${\bf f}_t$ has two components in spin space  which lay in the z-y plane
\begin{equation}
 \label{f-triplet-example}
 {\bf f}_t = \fpar \hat{\bf z} + f_t^{\perp}\hat{\bf y}\; .
\end{equation}
Now  the   system of Eqs.~(\ref{cov-Usadel-gen-s})-(\ref{cov-Usadel-gen-t}) reads
\begin{eqnarray}
 \label{Usadel-s-example}
 &&D\partial_x^2f_s - 2{|\omega|}f_s - 2i\sgn(\omega) {h} \fpar=0\; , \\
 \label{Usadel-par-example}
 &&D\partial_x^2\fpar + 2D\alpha\partial_xf_t^{\perp} 
 - 2\left(D\alpha^2 + {|\omega|}\right)\fpar - 2i\sgn(\omega) {h} f_s=0\; , \\
 \label{Usadel-perp-example}
&&D\partial_x^2 f_t^{\perp} -2D\alpha\partial_x\fpar 
 - 2\left(D\alpha^2 + {|\omega|}\right) f_t^{\perp}=0\; .
 \end{eqnarray}
Equations (\ref{Usadel-s-example}-\ref{Usadel-par-example}) describe diffusion of strongly coupled singlet and parallel triplet condensates. The last, singlet-triplet coupling, terms $\sim h$ in these equations dominate, and, as a result, both $f_s$ and $\fpar$ decay over the short length scale $\sim\xi_h$. Equation (\ref{Usadel-perp-example}) determines the spatial distribution of the perpendicular to ${\bf h}$ component $f_t^{\perp}$ of the triplet condensate. This component is generated near the interface because of the spin precession described by the second term and, according to Eq.~(\ref{Usadel-perp-example}),  decays over a much longer length scale . The spatial distribution of all components of the condensate is shown in  Fig. \ref{fig_sol}.

% %%%%%%%%%%%%%%%%%%%%%%%%%%%%%%%%%%%%%%%%%%%%%%%%%%%%%%%%%%%%%%%%%%%
%%%%%%%%%%%%%%%%%%%%%%%%%%
%%%%%%%%%%%%%%%%%%%%%%%%%%%%%%%%%%%%%%%%%%%
%
For a general SO field $\hat\Acal_k$ the LRTC is always induced if $\hat\Acal_k$ does not commute with the exchange field $\hat{h}$ and has a spatial component along the spin inhomogeneity. The condition $[\hat{h},\hat\Acal_k]\ne 0$ has an interesting interpretation in terms of SU(2) field tensor.\cite{BT2013} The exchange field enters the general many-body Hamiltonian as the time component of the SU(2) four-potential, 
$\hat{h}=\hat\Acal_0=\frac{1}{2}\Acal_0^a\hat\sigma^a$.\cite{Frolich93,Jin2006,Tokatly2008,Gorini2010} For the spatially uniform SU(2) potentials the above commutator is nothing but the SU(2) electric field 
$\hat\Fcal_{k0}=-i[\hat\Acal_k,\hat\Acal_0]$. Therefore the SU(2) electric field serves as a physical source of the LRTC in S/F structures, as it has been noticed recently in Ref.~\cite{BT2013}. We will return to this point in the next subsection.

At this stage it is important to emphasize the difference between the SO coupling studied here, originated from the band structure or geometrical constraints (such as hetero-interfaces),  and the SO effect caused by randomly distributed  impurities.\cite{Abrikosv_Gorkov1961} The latter case has been studied intensively in the context of S/F structures.\cite{Demler,Buzdin,BVE_SO} The only effect of the random SO coupling due to impurities is a finite, but fully isotropic spin relaxation rate. The direction of spin is always preserved and therefore no LRTC can be induced in this case.

We next show the   connection between the  inhomogeneous exchange field  and the SO coupling as sources of long-range triplet component.

\subsection{ LRTC in a Bloch-like domain wall: a gauge-equivalent interpretation}
\label{sec-diffS-Bloch}

The first theoretical work on the singlet-triplet conversion considered the case of a ferromagnet with a Bloch domain wall  at the interface with a superconductor.\cite{Bergeret2001}
It was assumed that the exchange field  ${\bf h}$ in the F layer of the inset of Fig. \ref{fig_sol}  follows  the magnetization direction that lies in the $y-z$ plane and rotates with respect to the $x$-axis.  Thus ${\hat h}$ in Eq. (\ref{Usadel-gen})  has the form:
\[
\hat h=h\left[\sin (Qx)\sigma^y+\cos( Qx)\sigma^z \right]\; ,
\]
where $Q$ is the wave-vector of the rotation.  In order to solve the  linearized Usadel equation it is convenient to  introduce the following local SU(2) rotation, as  done in Ref. \cite{Bergeret2001},
\begin{equation}
\label{rot_f}
\hat{\tilde f}(x)=U(x)\hat fU^{-1}(x)\; ,
\end{equation}
where $U(x)=e^{-\frac{i}{2}Qx\hat\sigma^x}$. Substitution of this expression into Eq.~(\ref{Usadel-gen}) removes the coordinate dependence from $h$
\begin{equation}
\label{rot-eq}
D\partial_{x}^2\hat{\tilde f}+\frac{DQ^2}{2}\left(\hat\sigma^x\hat{\tilde f}\hat\sigma^x-\hat{\tilde f}\right)+iDQ\left[\hat \sigma^x,\partial_x\hat{\tilde f}\right]-2|\omega|\hat{\tilde f}-i{\rm sgn}\omega\left\{h\hat\sigma^z,\hat{\tilde f}\right\}=0\; .
\end{equation}
One can easily  verify that this equation can be compactly written as:
\begin{equation}
\label{rot-cov}
D\tilde\nabla_{x}^2\hat{\tilde f}-2|\omega|\hat{\tilde f}-i{\rm sgn}\omega \left\{h\hat\sigma^z,\hat{\tilde f}\right\}=0\,  ,
\end{equation}
where 
 \begin{equation}
 \tilde\nabla_x\cdot=\partial_x\cdot + i\frac{Q}{2}\left[\hat\sigma^x,\cdot\right]\; .
 \end{equation}
Equation (\ref{rot-cov}) is identical to Eq. (\ref{cov-Usadel-gen}) for a homogeneous exchange field ${\bf h}=h\hat{\bf z}$ and a SO coupling described by a ``pure gauge'' SU(2) potential $\hat\Acal_x=-(Q/2)\hat\sigma^x$ and $\hat\Acal_y=\hat\Acal_z=0$. This a remarkable result that demonstrates that the problem of the singlet-triplet conversion in a S/F structure with a Bloch domain wall is gauge-equivalent to the one of a ferromagnet with an homogeneous exchange field and a SO coupling.  If we now compare Eq. (\ref{rot-eq}) with Eqs. (\ref{cov-Usadel-gen-s}-\ref{cov-Usadel-gen-t})  in the context of  the discussions in  sections \ref{sec-diffN} and \ref{sec-diffS-SO},  the second term in the l.h.s of Eq. (\ref{rot-eq}) describes the Dyakonov-Perel relaxation with anisotropy typical for a pure gauge SO coupling \cite{Tokatly_Ann}, while the third term induces the precession of the triplet component of the condensate and leads to the LRTC and the long-range proximity effect.  
 
This example  clearly shows the close connection between inhomogeneous exchange field and SO coupling by the generation of the LRTC. The inhomogeneous exchange field (inhomogeneous time component of the SU(2) potential $\hat\Acal_0$ ) at zero SO coupling $\hat\Acal_k=0$, and the homogeneous $\hat\Acal_0$ at nonzero $\hat\Acal_x$ describe the same physics in different gauges. The SU(2) electric field $\hat\Fcal_{k0} =\partial_k\hat\Acal_0 -i[\hat\Acal_k,\hat\Acal_0]$, which is the source of the LRTC, is present in both cases,  as it is a gauge covariant object. However, in one gauge $\hat\Fcal_{k0}\ne 0$ because of inhomogeneity of $\hat\Acal_0$, while in the other gauge due to nonvanishing commutator $[\hat\Acal_k,\hat\Acal_0]\ne 0$.
 
Before analyzing different S/F structures in the light of the SO coupling we present in the next section a more rigorous derivation of the main equation (\ref {cov-Usadel-gen}). Those readers  not interested in technical details can skip next section and go directly to section \ref{sec-examples} where we discuss the creation of long-range triplet correlations in different experimental setups.

\section{ Quasiclassical equations for systems with superconducting correlations, exchange field and spin-orbit coupling}\label{sec-equations}
 
The results  of the previous section are based on Eq. (\ref {cov-Usadel-gen}) which has been obtained by simple gauge invariance arguments. In this section we present a formal derivation of the  equations of motion for quasiclassical  Green's functions (GFs). We do not restrict our derivation to  the equilibrium case and introduce the 8$\times$8 Keldysh GFs matrix
\begin{equation}
\Gbb(\rv_1,\rv_2;t,t')=\left( \begin{array}{cc}
\check G^R&\check G^K\\
0&\check G^A
\end{array}
\right)  \label{GFK}
\end{equation}
where the retarded $\check{G}^R$, advanced $\check{G}^A$, and Keldysh $\check{G}^K$ GFs are 4$\times$4 matrices in the Nambu-Spin space. In principle we follow the standard derivation of the quasiclassical equation,\cite{LObook,BergeretRMP} but add the SO coupling described by the Hamiltonian (\ref{H-linearSO}). That is, we assume that SO coupling is linear in momentum, and the exchange field, $\hat{h}=h^a\sigma^a\equiv \hat\Acal_0$, does not depend on the momentum.  In such a case  the matrix (\ref{GFK}) obeys the Dyson equation
\begin{equation}
\left[i\tau_3\partial_t+\check G_0^{-1}+\check\Delta-{\Sigmabb}\right]\Gbb=1
\label{dyson1}
\end{equation}
where $\tau_3$ is the third Pauli matrix in Nambu space,
\[ 
\check G_0^{-1}=\tau_3 \hat{h}- \frac{1}{2m}\left(i\partial_k+\hat{\Acal}_k\right)^2+\mu\; ,
\]
\[
\check\Delta=\left( \begin{array}{cc}
0&\Delta\\
-\Delta^*&0
\end{array}
\right)\; ,
\]
$\mu$ is the chemical potential, $\Delta$ is the BCS order parameter and $\Sigmabb$ is the self-energy describing the elastic scattering at non-magnetic impurities. In the Born approximation the self-energy reads ${\Sigmabb}=(-i/2\tau)\langle{\gbb}\rangle$. Here $\tau$ is the elastic scattering time, ${\gbb}$ is the GF matrix integrated over quasiparticle energy and $\langle\dots\rangle$ stands for the average over the Fermi momentum direction. 

To simplify the derivation of the quasiclassical equations we assume for a moment that the exchange field $\hat{h}$ and the SO field $\hat\Acal_k$ do not depend on spatial coordinates. We will see that the full space dependence can be recovered at the end in the final equations from  symmetry arguments. 

By following  the standard route\cite{LObook} we first subtract from Eq. (\ref{dyson1}) its conjugate, and go to the Wigner representation in space by performing the Fourier transformation with respect to the coordinate difference $\xiv=\rv_1-\rv_2$. Then we proceed to the gradient expansion up to first order in derivatives with respect to the ``center of mass'' coordinate $\rv=\frac{\rv_1+\rv_2}{2}$. This procedure leads to the  following equation for the Wigner transformed matrix $\Gbb(\pv,\rv;t,t')$
\begin{equation}
\tau_3\partial_t{\Gbb}+\partial_{t'}{\Gbb}\tau_3-i\left[\tau_3\hat{h},{\Gbb}\right]+\frac{1}{2m}\left\{p_k-\hat{\Acal}_k, \partial_{k}{\Gbb}\right\}-i\frac{p_k}{m}\left[\hat{\Acal}_k, {\Gbb}\right]-i\left[\check\Delta,{\Gbb}\right]=-\frac{1}{2\tau}\left[\langle\gbb\rangle,{\Gbb}\right]\; .
\label{eqGorkov0}
\end{equation}
It is instructive to estimate the order of magnitude of different terms in this equation. Let $T$ and $L$ be characteristic time and length scales, that is, $\partial_t\sim 1/T$ and $\partial_{\rv}\sim 1/L$. Since $\Gbb$ as a function of $\pv$ is peaked at $p_F$, all momenta in Eq.~(\ref{eqGorkov0}) are of the order of $p_F$. Within the validity of semiclassical approach we assume that energies corresponding to $T^{-1}$, $v_F/L$, the momentum relaxation rate $\tau^{-1}$, the exchange energy $h$, SO spin splitting $v_F\Acal$, and the superconducting gap $\Delta$ are allowed to be of the same order of magnitude, but should all be much smaller than the Fermi energy $\varepsilon_F$. The ratio $\eta=\varepsilon_{\rm dyn}/\varepsilon_F$ of the above small dynamical energy scales to $\varepsilon_F$ is the small parameter that justifies the quasiclassical approximation in quantum kinetics. 

Now we can look on Eq.~(\ref{eqGorkov0}) from this point of view. Apparently all, except one, terms in Eq.~(\ref{eqGorkov0}) can be of the same order of magnitude being linear in the small parameter $\eta$. Only one term $\sim\hat{\Acal}_k\partial_{k}{\Gbb}$ in the left hand side has and extra factor of the order of $\Acal/p_F\sim\eta$. In the leading order of the semiclassical expansion it is absolutely natural to neglect this term. However it is also important to understand what are the physical consequences of this term and which effects we drop out by neglecting it. 

The physics of the subleading term can be revealed by transforming the kinetic equation Eq.~(\ref{eqGorkov0}) to the gauge covariant form in which SU(2) field strengths and the corresponding forces appear explicitly. For this sake we use the technique of gauge covariant Wigner functions, which has been developed originally in the context of quark-gluon kinetic theory\cite{Heinz} and applied more recently to describe the spin dynamics in semiconductors.\cite{Gorini2010} The main idea of this approach is to switch from the usual GF of Eq.~(\ref{GFK}) to its ``gauge covariant'' counterpart that is defined as follows
\begin{equation}
\label{tildeG} 
 \tilde{\Gbb}(\rv_1,\rv_2;t,t') = \hat{W}(\rv,\rv_1){\Gbb}(\rv_1,\rv_2,t,t')\hat{W}(\rv_2,\rv)\; ,
\end{equation}
where $\hat{W}(\rv,\rv_1)$ and $\hat{W}(\rv_2,\rv)$ are the Wilson link operators which ``covariantly connect'' the arguments of the Green's function to the "center-of-mass" coordinate  $\rv=\frac{\rv_1+\rv_2}{2}$. Formally the Wilson link operator entering this equation  is defined  by the  path-ordered exponential $\hat{W}(\rv_2,\rv_1) = P\exp{[i\int_{C_{12}}\hat{\Acal}_jdx_j]}$, where the integration path $C_{12}$ goes from $\rv_1$ to $\rv_2$ along the straight line. \cite{Heinz} The advantage of the GF $\tilde{\Gbb}$ in Eq.~(\ref{tildeG}) over the usual GF $\Gbb$ is that the Wigner transform of $\tilde{\Gbb}$,  and thus the corresponding quasiclassical GF $\tilde{\gbb}(\rv)$,  will transform locally covariantly under a nonuniform SU(2) rotation $\hat{U}(\rv)$, i.~e. $\tilde{\gbb}(\rv) \mapsto\hat{U}(\rv){\tilde \gbb}(\rv)\hat{U}^{-1}(\rv)$.  

In our case of spatially homogeneous SU(2) potentials $\hat{\Acal}_k$ the Wilson link operators reduce to a simple matrix exponential
$$
\hat{W}(\rv,\rv_1)=\hat{W}(\rv_2,\rv) = e^{\frac{i}{2}\hat{\Acal}_k(r_1^k-r_2^k)}.
$$
Obviously,  in this case the Wigner transformation of Eq.~(\ref{tildeG}) can be performed explicitly. The result is the following relation between the  Wigner transforms of the usual $\Gbb$ and the  covariant one  $\tilde{\Gbb}$ 
\begin{equation}
{\Gbb}({\bf r},{\bf p};t,t')=e^{-\frac{1}{2}{\hat\Acal}_k\overrightarrow{\partial_{p_k}}} \tilde{\Gbb}({\bf r},{\bf p};t,t') 
e^{-\frac{1}{2}{\hat\Acal}_k\overleftarrow{\partial_{p_k}}},
\label{tildeG-Wigner}
\end{equation}
where the upper arrow in the operators $\overrightarrow{\partial_{p_k}}$ and $\overleftarrow{\partial_{p_k}}$ indicate the direction in which the momentum derivative is acting.
% 
% Within the quasiclassical approach ${\Gbb}({\bf r},{\bf p};t,t')$ is given by 
% \begin{equation}
% {\Gbb}({\bf r},{\bf p};t,t')\approx \tilde{\Gbb}({\bf r},{\bf p};t,t') -\frac{1}{2}\left\{\hat\Acal_k,\partial_{p_k}\tilde{\Gbb}({\bf r},{\bf p};t,t')\right\}
% \end{equation}
% 

Now we can derive the equation for $\tilde{\Gbb}$ by substituting Eq.~(\ref{tildeG-Wigner}) into Eq. (\ref{eqGorkov0}) and then acting from the left with $\exp\{-\frac{1}{2}{\hat\Acal}_k\overrightarrow{\partial_{p_k}}\}$, and from the right with $\exp\{-\frac{1}{2}{\hat\Acal}_k\overrightarrow{\partial_{p_k}}\}$. Finally, by  making an expansion up to first order in the gradients and second order in SO fields $\hat{\Acal}_k$, we obtain the following equation for the gauge covariant function $\tilde{\Gbb}$ 
\begin{equation}
\label{eqGorkov-cv}
\tau_3\partial_t\tilde{\Gbb}+\partial_{t'}\tilde{\Gbb}\tau_3-i\left[\tau_3\hat{h},\tilde{\Gbb}\right]+ \frac{p_k}{m}\tilde{\nabla}_k\tilde{\Gbb} + 
\frac{1}{2\tau}\left[\langle\tilde{\gbb}\rangle,\tilde{\Gbb}\right]
-i\left[\check\Delta,\tilde{\Gbb}\right]= \frac{1}{2}\left\{\tau_3\hat\Fcal_{0k} 
+\frac{p_k}{m}\hat\Fcal_{kj},\partial_{p_j}\tilde{\Gbb}\right\}\; ,
\end{equation}
where $\tilde{\nabla}_k\cdot=\partial_k\cdot-i[\hat\Acal_k,\cdot]$ is the covariant gradient. In the right hand side of this equation we introduced the SU(2) field strength tensors $\hat\Fcal_{0k}=-i[\hat{h},\hat\Acal_k]$ and $\hat\Fcal_{kj}=-i[\hat\Acal_k,\hat\Acal_j]$.

Formally Eq.~(\ref{eqGorkov-cv}) was derived for spatially homogeneous exchange $\hat{h}$ and SO $\Acal_k$ fields. It is, however, absolutely clear that all we need to account for possible (static) inhomogeneities of the spin-dependent fields is to use for the SU(2) electric $\hat\Fcal_{0k}$ and $\hat\Fcal_{kj}$ the full expressions
\begin{eqnarray}
 \label{F0k}
\hat\Fcal_{0k}&=&-\partial_k\hat{h}-i[\hat{h},\hat\Acal_k],\\
\label{Fkj}
\hat\Fcal_{kj}&=&\partial_k\hat\Acal_j-\partial_j\hat\Acal_k-i[\hat\Acal_k,\hat\Acal_j].
\end{eqnarray}

An advantage of Eq.~(\ref{eqGorkov-cv}) over the original and more common Eq.~(\ref{eqGorkov0}) is the explicit SU(2) gauge covariance of the former. The SO coupling enters Eq.~(\ref{eqGorkov-cv}) only via the covariant gradient $\tilde{\nabla}_k$ and the SU(2) field tensor $\hat\Fcal_{\mu\nu}$. Now the physical significance of the subleading contribution to the kinetic equation can be easily identified. The subleading terms, of the order of $\eta^2$, are collected on the right hand side of Eq.~(\ref{eqGorkov-cv}). These terms describe the SU(2) Lorentz force \cite{Gorini2010} which, in particular, is responsible for the coupling of spin and charge degrees of freedom and the spin Hall effect. The leading contribution of SO coupling is exhausted by the covariant gradient term in the left hand side of Eq.~(\ref{eqGorkov-cv}). Physically it describes spin precession in the presence of the effective momentum dependent SO Zeeman field.

In the present paper we consider only the leading (spin precession) effects of SO coupling, while the terms of higher order in $\eta$ (the SU(2) Lorentz force effects) will be neglected. Obviously the latter have to be taken into account to study phenomena involving spin-charge coupling due to SO coupling.\cite{Stanescu2007,Gorini2010} It is worth noting that Eqs.~(\ref{eqGorkov0}) and (\ref{eqGorkov-cv}) become identical if we neglect terms of the order $\eta^2$.  

After neglecting the right hand side in Eq.~(\ref{eqGorkov-cv}) one can easily integrate this equation over the quasiparticle energy and, by using the fact that the ${\tilde\Gbb}$ is peaked at the Fermi level,  one obtains the SU(2) covariant Eilenberger equation
 \begin{equation}
\label{eqEilenberger}
\tau_3\partial_t{\gbb}+\partial_{t'}{\gbb}\tau_3-i\left[\tau_3\hat{h},{\gbb}\right] +v_Fn_k\tilde{\nabla}_k{\gbb}
-i\left[\check\Delta,{\gbb}\right]= -\frac{1}{2\tau}\left[\langle{\gbb}\rangle,{\gbb}\right]\; ,
\end{equation}
where ${\gbb}(\nv,\rv,t,t')$ is the quasiclassical GFs that depends on the Fermi momentum direction $\nv=\pv_{F}/p_F$, the center of mass coordinate ${\bf r}$ and two times.  In the diffusive case this equation can be further simplified by assuming that the GFs have a weak dependence on the momentum direction, {\it i.~e.} by approximating ${\gbb}\approx{\gbb_0}+{\nv}{\gbb_1}$. Following the standard derivation for diffusive equations (see for example \cite{LObook}) one finally arrives at the Usadel equation for the isotropic part ${\gbb_0}$ (we skip the index $0$):
 \begin{equation}
\label{eqUsadel}
D\tilde{\nabla}_k({\gbb}\tilde{\nabla}_k{\gbb})+ \tau_3\partial_t{\gbb} 
+ \partial_{t'}{\gbb}\tau_3-i\left[\tau_3\hat{h},{\gbb}\right]
-i\left[\check\Delta,{\gbb}\right]= 0,
\end{equation}
where $D=v_F^2\tau/3$ is the diffusion coefficient. We note that in the absence of superconducting correlations this equation leads to the spin diffusion equation (\ref{diffusion0}) or, equivalently, Eq.~(\ref{diffusion1}).
 
Throughout this article we only analyze equilibrium situations. In this case it is convenient to work with the Matsubara GF $\check{g}({\bf r},\omega)$ which is a $4\times 4$ matrix in the Nambu-spin space. The corresponding Usadel equation can be obtained straightforwardly from  Eq.~\ref{eqUsadel} (see for example \cite{BergeretRMP}):
 \begin{equation}
\label{eqUsadelM}
D\tilde{\nabla}_k(\check{g}\tilde{\nabla}_k\check{g})+ \omega[\tau_3,\check{g}]-i\left[\tau_3\hat{h},\check{g}\right]
-i\left[\check\Delta,\check{g}\right]= 0\; ,
\end{equation}
where $\omega$ is the Matsubara frequency.
Moreover,  we only focus on the linearized Usadel equation  which is valid either at temperatures close to the critical temperature or in the non-superconducting regions if the proximity effect is weak enough.  In such a case one can expand the GF's functions according to  $\check{g}\approx\tau_3{\rm sgn}(\omega)+i\tau_2{\hat f}$ where ${\hat f}$ is the anomalous Green function describing the superconducting condensate. We finally obtain
  \begin{equation}
\label{eqUsadelM-lin}
 D \tilde{\nabla}^2\hat{f} - 2|\omega|\hat{f}-i\sgn (\omega)\left\{\hat{h},\hat{f}\right\}= 0\; ,
\end{equation}
which coincides with  Eq.~(\ref{cov-Usadel-gen}) used along the manuscript.
%%%%%%%%%%%%%%%%
%%%%%%%%%%%%%%%%

\section{ Examples of singlet-triplet conversion in hybrid structures with SO coupling}
 \label{sec-examples}
  %%%%%%%%%%%%%%%%%%%%%%%%%%%%%%%%%%%%%%%%%%%%%%%%%%%%%%%%%%%%%%%%%%%
\begin{figure}[t]
\begin{center}
\includegraphics[scale=0.5]{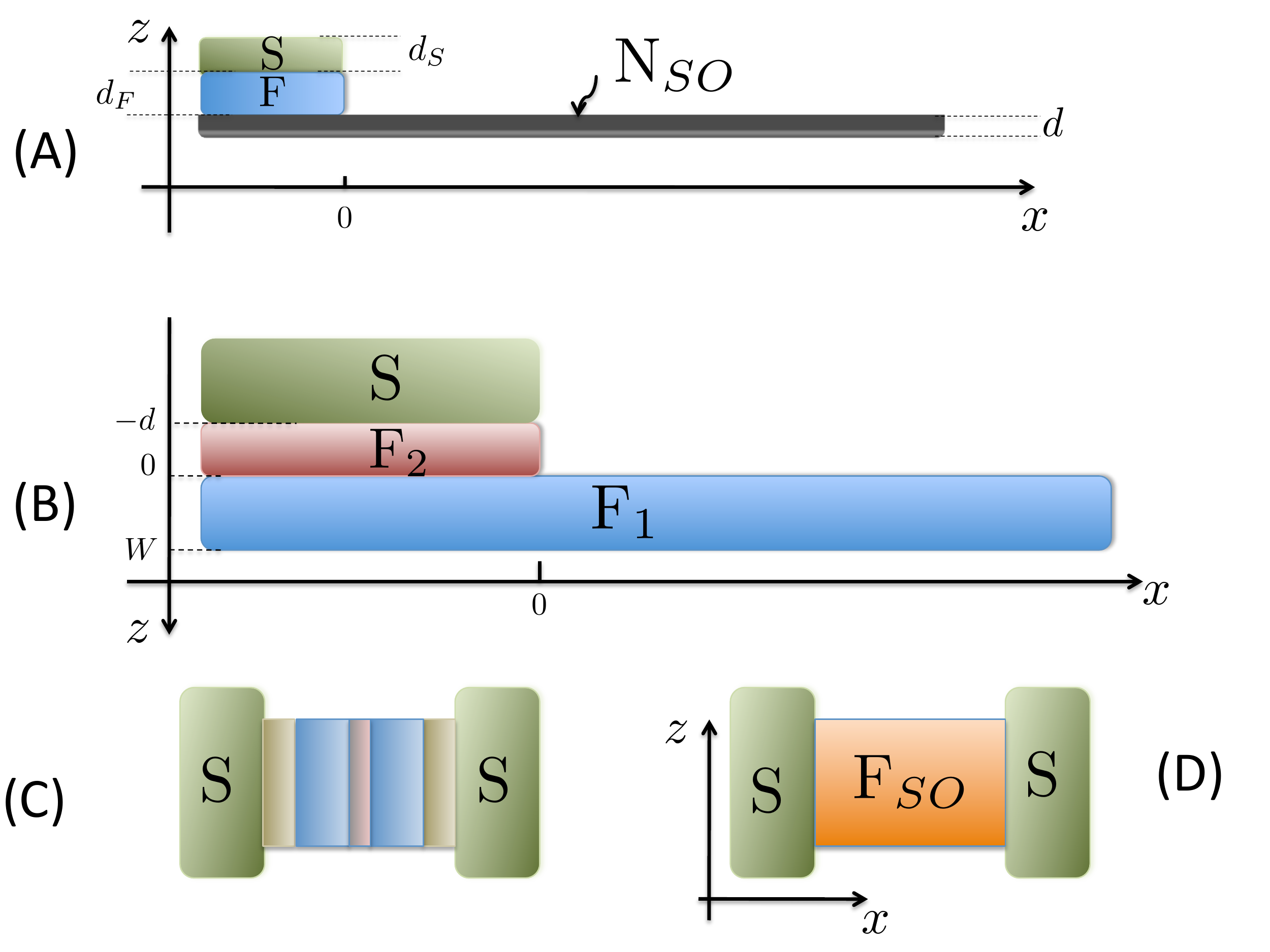}
\caption{\label{fig-geometries}Different geometries discussed in the main text. (A) A S/F/N$_{SO}$ structure. It is assumed that a finite Rashba SO coupling is present in the normal wire N$_{SO}$ . (B) A lateral S/F structure consisting of a thin ferromagnetic layer  F$_1$, a superconductor electrode and a second ferromagnetic layer,  F$_2$  between the S and the F$_1$. The S/F$_2$ structure extends over the $x<0 $ region. (C) Sketched of a transversal multilayer structure commonly used in  experiments and (D) its analog analyzed in the text. }
\end{center}
\end{figure}
%%%%%%%%%%%%%%%%%%%%%%%%%
%%%%%%%%%%%%%%%%%%%%%%%%%%%%%%%%%%%%%%%%%%
%%%%%%%%%%%%%%%%%%%%%%%%%%%%%%%%%%%%%%%%%%%%%%%%%%%%%%%%%%%%%%%%%%%
\subsection{ S/F/N structure with SO coupling}
As we have seen in sections \ref{sec-diffN} and \ref{sec-diffS-SO},  the SO coupling causes  both  the  spin rotation  in normal metals and the "rotation" of  the triplet  component of the condensate in S/F structures. From this analogy one  can infer  that if a triplet component is induced in a diffusive normal metal with SO coupling, such component may rotate leading to components perpendicular to the original one.  One can corroborate this statement by the following example, that represents a novel way of generation of the LRTC.

We consider  a  S/F/N$_{\rm SO}$  lateral structure as the one shown in Fig.~\ref{fig-geometries}A:  A S/F bilayer is situated on  top of a thin and narrow normal region, like a normal wire.\cite{jain0} The S/F bilayer extends to the left ($x<0$) and the normal wire has some intrinsic SO coupling. The F layer is sufficiently thin in order to allow for superconducting correlations to penetrate into the N wire. Notice that this geometry, without the F layer,  resembles pretty much the setup proposed for detection of Majorana fermions in hybrid structures. \cite{DasSarma,Majorana_exp} 

To simplify  formalities we assume that the S/F interface is transparent and  both layers are thin enough to describe them as an effective ferromagnetic superconductor\cite{Bergeret2001b,Giazotto2013} with effective values for the order parameter $\Delta_{eff}=\Delta\nu_Sd_S/(\nu_Sd_S+\nu_Fd_F)$ and the exchange field $h_{eff}=h\nu_Fd_F/(\nu_Sd_S+\nu_Fd_F)$, where $d_{S(F)}$ is the thickness of the S(F) layer and $\nu_{S(F)}$ its density of the states. Thus, the SF layer  exhibits a BCS-like density of states which is now spin-dependent shifted by $h_{eff}$. If the exchange field lies in the $(x,y)$ plane (see  Fig. \ref{fig-geometries}A)  the condensate function  in the S/F electrode consists, as usual, of a singlet $f_{FS}^s=f_+$ and a triplet  component that reads
\begin{equation}
\hat f_{SF}^t=f_-(\cos\theta\sigma^x+\sin\theta\sigma^y)\; ,
\end{equation}
where $f_{\pm}=[f_{BCS}(\omega+ih_{eff})\pm f_{BCS}(\omega-ih_{eff})]/2$, and $\theta$ is the angle between the exchange field and the $x$-axis. In this way, the function of the S/F electrode is to generate in the normal metal  the triplet component parallel to the exchange field of F.  In analogy to the spin diffusion   in a normal metal ({\it cf.} Section \ref{sec-diffN}), the induced triplet component is eventually rotated   in the N$_{\rm SO}$ wire  and all other triplet components generated as we discuss next.   

\begin{figure}[h]
\begin{center}
\includegraphics[width=\columnwidth]{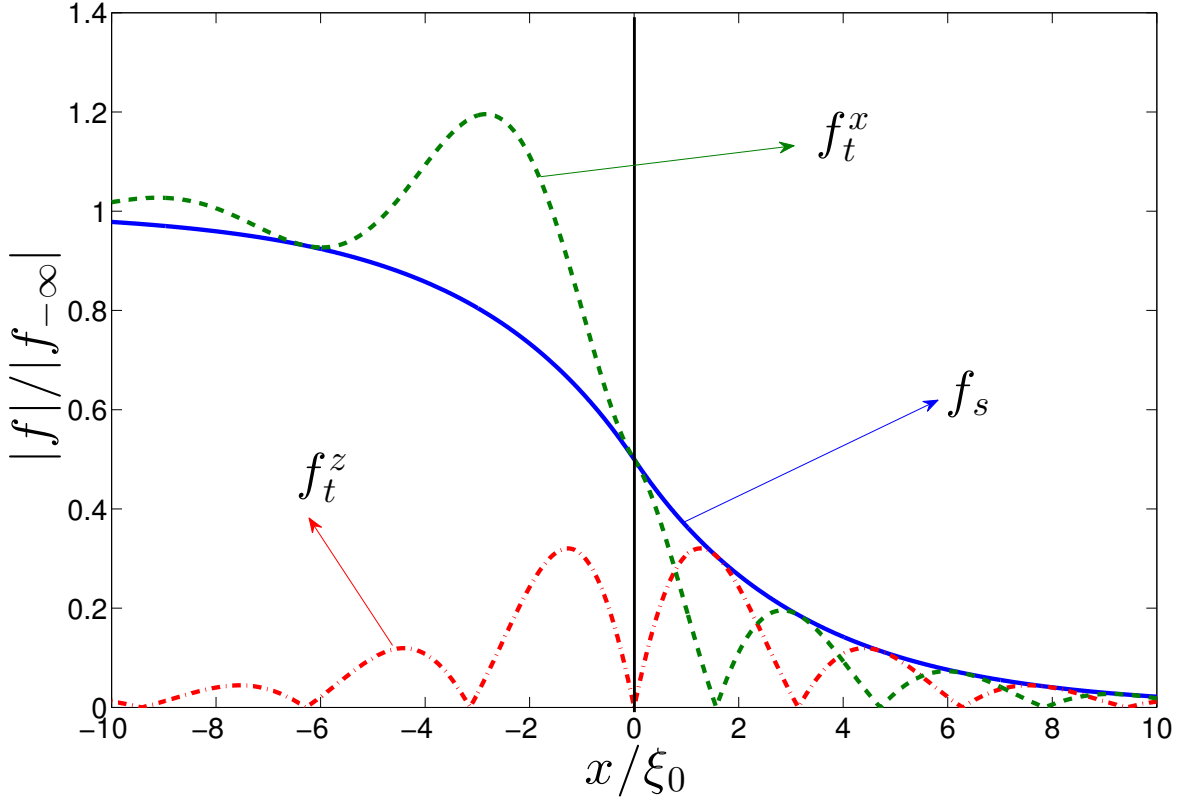}
\caption{\label{fig_SFN}The spatial behavior of the singlet   and  triplet   components $f_t^z$ and $f_t^x$  in the normal region of the structure shown in Fig. \ref{fig-geometries}A obtained from Eqs.~(\ref{eqs-SFN1})-(\ref{eqs-SFN4}).  We have chosen $\omega=\pi T$,  $\alpha\xi_0=1$ and $T=0.1\Delta_0$, and $\theta=0$. The values of $f_s$ and $f_t^x$ are normalized to their asymptotic values at $x=-\infty$, while $f_t^z$ is normalize to the asymptotic value of $f_t^x$.}
\end{center}
\end{figure}

If the  N$_{\rm SO}$ wire is deposited on a substrate it is natural to assume that the  SO coupling is described by  $\Acal_x^y=\alpha$, while  all other components of $\hat\Acal$ are zero.  Moreover,  we assume that the  width $d$ of the normal wire is much smaller than the characteristic variation of condensate induced via proximity effect. Thus, we can integrate the Usadel Eq. (\ref{cov-Usadel-gen}) over the z-direction by using the boundary condition Eq. (\ref{cov-BC-gen}) which now reads:
\begin{eqnarray}
\partial_z f_s&=&-\gamma f_+^s\\
\partial_zf_t^a\sigma^a&=&-\gamma f_-\left(\cos\theta\delta^{ax}+\sin\theta\delta^{ay}\right)\;.
\end{eqnarray}
With all these assumptions and after integration over $z$-direction we  end up with the following set of 1D  linear differential equations:
\begin{eqnarray}
\label{eqs-SFN1}
\partial^2_xf_s-\kappa_\omega^2f_s&=&\frac{\gamma}{d}f_+\Theta(-x)\\
\partial^2_xf_t^y-\kappa_\omega^2f_t^y&=&\frac{\gamma}{d}f_-\sin\theta\Theta(-x)\\
\partial^2_xf_t^x-(\kappa_\omega^2+\alpha^2)f_t^x+2\alpha\partial_xf_t^z&=&\frac{\gamma}{d}f_-\cos\theta\Theta(-x)\\
\label{eqs-SFN4}
\partial^2_xf_t^z-(\kappa_\omega^2+\alpha^2)f_t^z-2\alpha\partial_xf_t^x&=&0\; ,
\end{eqnarray}
where $\Theta(x)$ is the Heaviside step-function and $\kappa_\omega^2=2|\omega|$.  It is straightforward to obtain the solution of this system. We present here only the solution for the triplet components of the condensate in the region $x>0$ (Fig.~\ref{fig-geometries}A):
\begin{equation}
\label{triplet-SFN}
\hat f_t={\cal C}_\omega f_-e^{-\kappa_\omega x}\left[\cos(\alpha x)\cos\theta\sigma^x+\sin\theta\sigma^y+\sin(\alpha x)\cos\theta\sigma^z\right]\; ,
\end{equation}
where ${\cal C}_\omega=-\gamma/[2d(\kappa_\omega^2+\alpha^2)]$. As expected,  
the "injected" triplet component  of the condensate, which is parallel to the exchange field of the S/F bilayer, can rotate if a finite  SO coupling exists in the normal region.  For this to occur  the  SO coupling must satisfy  $[\hat\Acal_k,\hat\Acal_0]\ne 0$. In our particular case ($\hat\Acal_x=\frac{1}{2}\alpha\sigma^y$)  the perpendicular components are generated provided that the exchange field is not pointing in the $y$-direction. In the latter case, as one can directly see from Eq.~(\ref{triplet-SFN}), only the parallel component is generated in N$_{\rm SO}$.   The presence of the SO coupling leads to a spatial  oscillation of the $f^x_t$ and $f_t^z$ components as shown in Fig.~\ref{fig_SFN}. We should emphasize however, that this oscillation has another origin as the one discussed in the context of SF structures without spin-orbit.\cite{BuzdinRMP,Ryazanov2001} In the latter case the oscillations in the F layer are due to the presence of a (homogenous) exchange field which also affects the singlet 
component.  Here however,  
there is no exchange field in the N region and the oscillations  are simple due to the SO term in analogy to the spin rotation in normal systems. Notice that in our geometry the singlet component does not oscillate and  no 0-$\pi$ transition is expected in a symmetric S/F/N$_{\rm SO}$/F/S Josephson junction, in contrast to the oscillations in the critical current observed in  SFS structures.\cite{Ryazanov2001} 

In principle the S/F/N$_{\rm SO}$ structure described here can be used as a generator of the LRTC . If we assume, for example that   at the other end of the N$_{\rm SO}$ wire there is second strong  ferromagnet with a magnetization parallel to the "injector" F the  component $f_t^x$ of the condensate will penetrate this second ferromagnet over long distances of the order of $\sqrt{D/T}$.  
We notice that the mechanism discussed in this section also explains the triplet component induced in a superconductor-2D normal metal-superconductor junction  with Rashba SO coupling in an external Zeeman field discussed in Ref.\cite{Malshukov}.  It is also worth to mention that a similar (but in the ballistic limit) has been  analyzed in a recent manuscript.\cite{jain0}
  %%%%%%%%%%%%%%%%%%%%%%%%%%%%%%%%%%%%%%%%%%%%%%%%%%%%%%%%%%%%%%%%%%%
%%%%%%%%%%%%%%%%%%%%%%%%%
%%%%%%%%%%%%%%%%%%%%%%%%%%%%%%%%%%%%%%%%%%
%%%%%%%%%%%%%%%%%%%%%%%%%%%%%%%%%%%%%%%%%%%%%%%%%%%%%%%%%%%%%%%%%%%

\subsection{Lateral Josephson junction with SO coupling}
The first evidence of  long-range superconducting correlations in magnetic materials was found  by measuring a finite supercurrent flowing through a half-metallic CrO$_2$ in a lateral Josephson junction \cite{Keizer2006}  (see sketch in Fig.~\ref{fig-geometries}B).   The  supercurrent across the junction could be observed  up to distances of the order  of one micron between the S leads and  can only be explained by assuming  that   the supercurrent is carried by Cooper pairs with equal spin projection, or in our terminology by assuming a finite triplet component of the condensate perpendicular to the magnetization direction of the half-metal.   The required    spin-triplet conversion  might take place in a region around the  S/F interface  if one  assumes a  magnetic disorder with a finite  averaged moment misaligned with respect to the bulk magnetization of the  CrO$_2$ layer.\cite{Eschrig2008}    It is difficult to prove experimentally such inhomogeneity. More recent experiments on  CrO$_2$ based Josephson 
junctions have shown that the observation of long-range effects depends on the substrate on which the half metal is grown.   For example, generation of the long-range triplet component has been observed in CrO$_2$   grown onto Al$_2$O$_3$ by using simple superconducting contacts. In contrast, if the  CrO$_2$ is grown onto a TiO$_2$ substrate, the long-range Josephson effect can only be observed  if one incorporates a thin Ni layer between the CrO$_2$  and the superconducting electrodes. \cite{Aarts2010,Aarts2012} It is commonly  believed that in both cases the long-range triplet component is generated due  to a magnetic inhomogeneity, either originated at the   superconductor/CrO$_2$ interface (spin-active interface) or in the Ni interlayer.\cite{Yates2013}

We give here an additional possible explanation for the long-range proximity effect in such lateral structures, based on the presence of SO coupling at the contact region. The existence of a SO coupling in the CrO$_2$ experiments was suggested in Ref.~\cite{Yates2013}, but not discussed quantitatively due to the lack of a formalism for this. 
We have now all ingredients to include the SO coupling in the study of the proximity effect, and focus our analysis on the system sketched in  Fig.~\ref{fig-geometries}B . It is a lateral structure consisting of a superconductor S and a ferromagnet F$_1$.  At the interface between them there is an additional thin layer, F$_2$,   with a magnetization parallel to the F$_1$ layer. Thus, in principle,  one does not expect any long-range effect in accordance with previous theories. \cite{BergeretRMP}
We assume that in F$_2$ there is a finite SO coupling , which can be either due to some crystallographic inversion asymmetry\cite{Samokhin2009} or due to the presence of interfaces between materials and the lack of structure inversion symmetry.\cite{Rashba,Edelstein2003,Koroteev2004,Ast2007,Miron2010,Linder2011,Duckheim2011,Takei2012,Takei2013} 

The S/F$_2$ bilayer extend over the whole negative $x-axis$ and the  SO coupling is only present in the F$_2$ layer and therefore the SO vector potential is written as:
\begin{equation}
\label{SO-lateral} 
\Acal^a_j(z,x)=\Acal^a_j\Theta(-x)\Theta(-z)\Theta(z+d)\; .
\end{equation}
If one assumes translation invariance in $y$ direction then the condensate function in Eq.~(\ref{cov-Usadel-gen}) depends  on $x$ and $z$ coordinates (see Fig.~\ref{fig-geometries}B) and satisfies Eqs. (\ref{cov-Usadel-gen-s}-\ref{cov-Usadel-gen-t}).
%  one should solve the  two dimensional boundary problem 
%\begin{equation}
%D\partial_k\partial_k \hat f-2iD\left[\hat \Acal_j,\partial_j \hat f\right]-D\left[\hat \Acal_j,\left[\hat\Acal_j,\hat f\right]\right]-2|\omega|\hat f-\frac{i{\rm sgn}\omega \; h}{2}\left\{{\bf m_0.\sigma},\hat f\right\}=0;\label{Usadel_lat}
%\end{equation}
%where ${\bf m_0}$ is a constant unit vector in the $x-y$ plane and 
%\begin{eqnarray}
%\Acal_j(z,x)&=&\Acal_j\Theta(-x)\Theta(-z)\Theta(z+d)\\
%h(z,x)&=&h_2\Theta(-x)\Theta(-z)\Theta(z+d)+h_1\Theta(z)\Theta(W-z)
%\end{eqnarray}
%with the Kupryianov-Lukichev boundary conditions at the S/F$_1$ and F$_1$/F interfaces.
This problem  can be solved numerically. However, in order to underline the physics of the singlet-triplet conversion we solve here the problem analytically by assuming first  that the total thickness $W+d$ is much smaller than the characteristic length over which the condensate $f$ changes.  This assumption  allows us to integrate the Usadel equation over $z$. Secondly,  we  neglect quadratic terms in $\Acal$,  by assuming that $|\Acal|^2\ll h/D $. This means we neglect the term proportional to $\Gamma$ in Eq. (\ref{cov-Usadel-gen-t}).
After integration over  the $z$-direction and by using the boundary condition Eq. (\ref{cov-BC-gen}) at the S/F$_2$ and continuity at F$_1$/F$_2$ interfaces  we obtain from Eqs. (\ref{cov-Usadel-gen-s}-\ref{cov-Usadel-gen-t})
 \begin{eqnarray}
D\partial ^2_{xx}fs-2|\omega|f_s-2i{\rm sgn}(\omega) h^xf^x_t &=&-\frac{D\gamma}{W+d} f_{BCS}\label{lat1xl0}\; ,\\
D\partial ^2_{xx}f_t^a-2|\omega|f_t^a-2i{\rm sgn}(\omega) h^xf_s+\frac{2d}{d+W}{C}^{ab}_x\partial_xf_t^b &=&0\; ,
\label{lat2xl0} 
\end{eqnarray}
for $x<0$,  and 
 \begin{eqnarray}
D\partial ^2_{xx}f_s-2|\omega|f_s-i{\rm sgn}(\omega) h^x_1f^x_t &=&0\; ,
\label{lat1xg0}\\
D\partial ^2_{xx}f_t^a-2|\omega|f_t^a-i{\rm sgn}(\omega) h^x_1f_s&=&0\; ,
\label{lat2xg0} 
\end{eqnarray}
for $x>0$. Here $h_{1,2}$ are the exchange fields (that point in $x$-direction) in the F$_{1,2}$ regions and $h=h_1W/(W+d)+h_2d /(W+d)$ is the averaged exchange field.  We assume that the SO coupling is of Rashba type with $C_x^{xz}=-D\alpha$. 
These equations have to be solved assuming that the condensate function is continuous at $x=0$ and 
\begin{equation}
\label{lat-BC}
\left.\partial_x \hat f+i\frac{d}{d+W}\alpha\left[\sigma^y,\hat f\right]\right|_{x=0^-}=\left.\partial_x \hat f\right|_{x=0^+}\; .
\end{equation}
From a simple inspection of  Eqs. (\ref{lat1xl0}-\ref{lat-BC})  one can conclude that a finite triplet component, $f_t^x$ perpendicular to the exchange field is generated by the SO coupling term. The decay of this component into the $x>0$ is 
long-range as follows from Eq. (\ref{lat2xg0}).    Deep in the region  covered by the S/F$_2$ bilayer ($x\rightarrow-\infty$)  the solution does not depend on $x$ and  according to  Eq. (\ref{lat1xl0}) is given by
\begin{eqnarray}
f_s(-\infty)&\approx& \frac{\gamma\xi_h^2}{W+d}\frac{|\omega|}{2h}f_{BCS}\\
f_t^x(-\infty)&\approx&-i \frac{\gamma{\rm sgn}(\omega)\xi_h^2}{2(W+d)}f_{BCS}\\
f_t^z(-\infty)&=&0;\, 
\label{asympt}
\end{eqnarray}
where $\xi_h=\sqrt{D/h}$ (we have assumed that $h_1\gg h_2, T$).   Notice that the asymptotic value of the "perpendicular" component of the triplet $f_t^z$,  is zero. 
In principle one can obtain straightforwardly the spatial dependence for all condense components by solving the boundary problem  Eqs. (\ref{lat1xl0}-\ref{lat-BC}). Here  we present the solution for the long-range  component $f_t^z$ in the region $x>0$. It is given by
\begin{equation}
f_t^z(x>0)=\frac{3}{2}i{\rm sgn}(\omega)\frac{\alpha d\xi_\omega\xi_h^2}{(W+d)^2}f_{BCS}e^{-\kappa_\omega x}  
\label{lat-LRTC} 
 \end{equation}
where $\kappa_\omega^{-1}=\sqrt{D/2|\omega|}$ is  the characteristic decay-length. 
If we now assume that at  $x=L\gg\xi_h$ there is a second S/F$_2$ electrode one can easily shown  the  (spectral) Josephson  current through the junction decays as \cite{BT2013} $e^{-\kappa_\omega L}$ with the junction length $L$. This  confirms the long-range character of the proximity effect. It is important to emphasize that the  magnetization of all F layers has been assumed to be parallel. The   long-range component,  Eq. (\ref{lat-LRTC}),  is proportional to  the SO coupling $\alpha$  in the F$_2$ thin layer. This example clearly shows that besides magnetic inhomogeneity, the SO coupling can also be a source for the LRTC. 

\subsection{Multilayer transversal structures }

Apart of the experiments on CrO$_2$ lateral structures, most of the experiments  searching for  triplet  long-range proximity effect have been performed on transversal structures\cite{Birge2010,Robinson2010,Sprungmann2010,Usman2011,Birge2012}  as the one sketched  in Fig. \ref{fig-geometries}C.  The region between the S electrodes consists of a multilayered magnetic structure that provides the magnetic inhomegeneity for the singlet-triplet conversion. 
Again, due to the hetero-interfaces between different materials one  can expect a finite  SO coupling in the structure.\cite{Edelstein2003,Linder2011}
In order to simplify the problem, instead of analyzing the multilayer system of Fig. \ref{fig-geometries}C, we study here the  SF$_{\rm SO}$S junction of Fig. \ref{fig-geometries}D, by assuming that the F$_{\rm SO}$, besides the in-plane exchange field, exhibits also a SO coupling of the form:
\begin{eqnarray}
\label{trans-SO-1}
\hat \Acal_z&=&\beta \sigma^z-\alpha\sigma^y\; ,\\
\label{trans-SO-2}
\hat \Acal_y&=&-\beta\sigma^y+\alpha\sigma^z\; ,
\end{eqnarray}
where $\alpha$ and $\beta$ are known in the literature as the  Rashba and 
Dresselhaus constants respectively. The system is translation invariant in the $(x,y)$ plane and therefore 
it is unlikely to have a finite  component of the vector potential $\hat\Acal_k$ in $z$ direction. Moreover the condensate function $\hat f$ only varies over $x$ direction and therefore the second term in Eq. (\ref{cov-Usadel-gen-t}) does not contribute. This means that, eventually, the only source for the LRTC is  the relaxation rate tensor $\Gamma^{ab}$, defined in Eqs. (\ref{diffusion1}-\ref{Gamma-linear}). Thus, the condition for generating the long triplet component, {\it i.~e.} the component perpendicular to the exchange field is that the vector $\left[\hat\Acal_k,\left[\hat\Acal_k,h^a\sigma^a\right]\right]$ {\it is not} parallel to the exchange field $h^a\sigma^a$.  For the SO coupling described by Eqs. (\ref{trans-SO-1}-\ref{trans-SO-2}) one obtains
\begin{equation}
\label{cond-triplet}
\left[\hat\Acal_k,\left[\hat\Acal_k,h^a\sigma^a\right]\right]=4(\alpha^2+\beta^2)(h^a\sigma^a+h^x\sigma^x)+8\alpha\beta(h^y\sigma^z+h^z\sigma^y)\; .
\end{equation}
If the magnetization points in the perpendicular direction ({\it i. e.} $h^y=h^z=0$)  then the LRTC is not generated.  If all components of the exchange field are finite (as in the case of Ho layers \cite{Robinson2010}) the term proportional to $\Gamma^{ab}$ in Eq. (\ref{cov-Usadel-gen-t}) generates LRTCs for any value of $\alpha$ and $\beta$.

In the most common case of an in-plane magnetization $h^x=0$, the condition for the  LRTC is  that $\alpha\beta\neq0$ and $h^y\neq h^z$. It is important to emphasize that this condition for triplet generation  is more  restrictive than in the lateral geometry studied in previous section, in which a pure Rashba SO coupling   at the S/F interface and arbitrary magnetization orientation  are enough for the LRTC to exist.

\section{Conclusions} 

The SO coupling discussed here has its  origin in the lack of inversion symmetry and therefore it has to be distinguished from the SO  coupling   originated by disorder which does not generate  the long-range triplet component and it was widely studied in the past decades. On the one hand the lack of inversion symmetry  can be due to  some crystallographic inversion asymmetry  in the materials. However, such noncentrosymmetric metals  have not been experimentally explored in the context of superconducting proximity effect. A detailed  analysis of these materials based on the symmetry arguments can be found in the review Ref. \cite{Samokhin2009} . On the other hand, the lack of inversion symmetry  can also occur at the interface between two different materials inducing an interfacial  SO coupling.\cite{Edelstein2003,Koroteev2004,Ast2007,Miron2010,Linder2011,Duckheim2011,Takei2012,Takei2013} This might be the scenario in some of the  structures used in the experiments on SFS junctions. It is not 
straightforward to 
estimate the strength of the SO coupling 
for a given hybrid interface. This  has  been obtained from first principle calculations for certain material combinations.\cite{Ast2007} 
Also experiments exploring spin torque  in Pt/Co/AlO$_x$  multilayer,  provide a pretty large value for the SO coupling induced by the inversion asymmetry of the structure.\cite{Miron2010}  A considerable SO coupling is also predicted for other metallic interfaces.\cite{Manchon2008}

In conclusion, we have presented an exhaustive study of the proximity effect in diffusive superconductor-ferromagnet hybrid structures  with spin-orbit coupling. We have derived the quasiclassical equations that include  generic spin fields. For the particular case of spin-orbit coupling  linear in momentum,  we have drawn an  useful and novel   analogy between the spin precession in a normal diffusive system with SO coupling and the generation of the long-range triplet component in S/F structures.  As for a spin density in a  normal system , the presence of a SO coupling may rotate the triplet component of the superconducting condensate and generate all triplet projections. We explicitly demonstrate that both,  the spin diffusion equation in the normal state and the linearized Usadel equation describing the proximity effect in SF structures with SO coupling,  are almost identically.  This analogy provides a useful tool for the design  of experimental setups and the search of optimal  material combinations 
for the control and  manipulation of the triplet component in hybrid superconducting structures. 
Moreover, it suggests a possible way to control and manipulate   the spin  in low dissipative devices based on S/F hybrids with spin-orbit coupling.  
 As an example of this, we have shown that a normal wire with an intrinsic SOC attached to a S/F electrode,  can be the  source for the long-range triplet component.
We also   predict  the appearance of long-range triplet in a variety of S/F diffusive systems in which the SO coupling is finite and demonstrate that the singlet-triplet conversion via SO coupling is more likely to happen in lateral structures rather then multilayer transversal systems. Our results  can be easily extended for arbitrary spin fields  and thus  unify in a natural  way all mechanisms  for the singlet-triplet conversion, providing  a useful tool for the description of the physics underlying superconducting hybrid systems with generic spin-fields.

\section {Acknowledgements}   We thank Vitaly Golovach for useful discussions. 
The work of F.S.B  was supported by  the Spanish Ministry of Economy and Competitiveness under
Project FIS2011-28851-C02-02  the Basque Government under UPV/EHU Project IT-756-13.  I.V.T. 
 acknowledges funding by the ``Grupos Consolidados UPV/EHU del Gobierno Vasco'' (IT-578-13) and Spanish MICINN (FIS2010-21282-C02-01). F.S.B thanks Martin Holthaus and his group for their kind
hospitality at the Physics Institute of the Oldenburg University.

\end{document}